%% file: main.tex
\pdfoutput=1
\documentclass[sigconf, authorversion]{acmart} 

\usepackage[utf8]{inputenc}
\usepackage[T1]{fontenc}
\usepackage{nameref}

\usepackage{hyphenat}

\usepackage[english]{babel}
\usepackage{centernot}
\usepackage{cancel}

\usepackage[ruled, vlined]{algorithm2e}
\SetKwComment{Comment}{$\triangleright$\ }{}
\SetCommentSty{itshape}

\usepackage[position=b]{subcaption}
\usepackage{xcolor}
\usepackage{ifthen}

\usepackage{tikz}
\usetikzlibrary{automata, positioning, arrows, arrows.meta, decorations.pathreplacing, calc, chains, shapes.misc, backgrounds, fit}
\usepackage{pgfplots}
\pgfplotsset{compat=newest}

\usepackage{hyperref}
\usepackage{multirow}
\usepackage{framed}

\pgfdeclarelayer{background}
\pgfsetlayers{background, main}

\usepackage[ruled]{algorithm2e}
\SetKwComment{Comment}{$\triangleright$\ }{}
\SetCommentSty{itshape}

\usepackage{listings}

\input{tikz-hypergraph}

\copyrightyear{2022} 
\acmYear{2022} 
\setcopyright{none} 
\acmConference[ESEM '22]{ACM/IEEE International Symposium on Empirical Software Engineering and Measurement (ESEM)}{September 19--23, 2022}{Helsinki, Finland}
\acmBooktitle{ACM/IEEE International Symposium on Empirical Software Engineering and Measurement (ESEM) (ESEM '22), September 19--23, 2022, Helsinki, Finland}
\acmDOI{10.1145/3544902.3546254}
\acmISBN{978-1-4503-9427-7/22/09}

\begin{document}

\title{Only Time Will Tell: Modelling Information Diffusion in Code Review with Time-Varying Hypergraphs}

\author{Michael Dorner}
\orcid{0000-0001-8879-6450}
\affiliation{%
	\institution{Blekinge Institute of Technology}
	\city{Karlskrona}
	\country{Sweden}
}
\email{michael.dorner@bth.se}

\author{Darja Šmite}
\orcid{0000-0003-1744-3118}
\affiliation{%
	\institution{Blekinge Institute of Technology}
	\city{Karlskrona}
	\country{Sweden}
}
\email{darja.smite@bth.se}

\author{Daniel Mendez}
\orcid{0000-0003-0619-6027}
\affiliation{%
	\institution{Blekinge Institute of Technology}
	\city{Karlskrona}
	\country{Sweden}
}
\affiliation{%
	\institution{fortiss GmbH}
	\city{Munich}
	\country{Germany}
}
\email{daniel.mendez@bth.se}

\author{Krzysztof Wnuk}
\orcid{0000-0003-3567-9300}
\affiliation{%
	\institution{Blekinge Institute of Technology}
	\city{Karlskrona}
	\country{Sweden}
}
\email{krzysztof.wnuk@bth.se}

\author{Jacek Czerwonka}
\orcid{0000-0002-4924-5047}
\affiliation{%
	\institution{Microsoft Research}
	\city{Redmond}
	\country{USA}
}
\email{jacekcz@microsoft.com}

\begin{abstract}
\noindent\textbf{Background:} %
Modern code review is expected to facilitate knowledge sharing: All relevant information, the collective expertise, and meta-information around the code change and its context become evident, transparent, and explicit in the corresponding code review discussion. The discussion participants can leverage this information in the following code reviews; the information diffuses through the communication network that emerges from code review. Traditional time-aggregated graphs fall short in rendering information diffusion as those models ignore the temporal order of the information exchange: Information can only be passed on if it is available in the first place.

\noindent\textbf{Aim:} %
This manuscript presents a novel model based on time-varying hypergraphs for rendering information diffusion that overcomes the inherent limitations of traditional, time-aggregated graph-based models. 

\noindent\textbf{Method:} %
In an in-silico experiment, we simulate an information diffusion within the internal code review at Microsoft and show the empirical impact of time on a key characteristic of information diffusion: the number of reachable participants. 

\noindent\textbf{Results:} %
Time-aggregation significantly overestimates the paths of information diffusion available in communication networks and, thus, is neither precise nor accurate for modelling and measuring the spread of information within communication networks that emerge from code review. 

\noindent\textbf{Conclusion:} %
Our model overcomes the inherent limitations of traditional, static or time-aggregated, graph-based communication models and sheds the first light on information diffusion through code review. We believe that our model can serve as a foundation for understanding, measuring, managing, and improving knowledge sharing in code review in particular and information diffusion in software engineering in general.
\end{abstract}



\maketitle

\section{Introduction}

Code review has transformed over the last decades from a waterfall-like procedure primarily used for detecting bugs in formal, heavyweight code inspections in the 1980s to a knowledge-sharing platform in an informal, tool-supported, lightweight process nowadays \cite{Bacchelli2013, Bosu2017, Baum20161}. Since modern software systems are often too large, too complex, and evolving too fast for a single developer to oversee all parts of the software and, therefore, to understand all implications of a change, most software projects use code review to foster a broad discussion on the change and its impact before it is merged into the code base. Each code review becomes a communication channel to share knowledge among the discussion participants: All relevant information, collective expertise, and meta-information about the change become evident, transparent, and explicit through those discussions and are shared among the participants. Since the participants implicitly cache this information, they can use, build upon, and spread it in the upcoming code reviews they participate in. Over time, information is spread through code review among its participants, the so-called \emph{information diffusion}.

Until today, software engineering research relies on time\hyp{}aggregated graph-based models for representing communication networks of all kinds~\cite{Herbold2021}. However, time-aggregated, graph-based communication models are not capable of rendering such an information diffusion since information diffusion is neither necessarily bilateral nor instant: In a discussion during a code review, multiple people can receive information concurrently and information can only be passed on if it is available to the participant beforehand. Depending on the temporal availability of vertices and edges, the information takes different routes through the communication network---a type of network topology traditional, time-aggregated graph-based communication models cannot render. 

Motivated by these shortcomings, we introduce a novel model for information diffusion in channel-based communication based on time-varying hypergraphs to research how information originated from a code review discussion spread among the communication participants. We validate our time-respecting model in comparison with an equivalent but a time-aggregated graph-based model in a computer simulation that empirically shows the impact and importance of time-awareness for information diffusion analysis in code review. For this comparison, we use a key characteristic for information diffusion, the number of reachable participants, which also reflects the number of paths in a communication network that are valid and available for direct and indirect information exchange.

The main contributions of this manuscript are as follows:

\begin{itemize}
\item We introduce a novel communication model based on time-varying hypergraphs for information diffusion within communication networks. 
\item To this end, we provide a concise and gentle introduction to the mathematical foundation of time-dependent hypergraphs and the impact of topological and temporal distance on information diffusion modelling. 
\item We simulate the spread of information within the communication network emerging from code review at Microsoft to validate our model compared to an equivalent but time-aggregated model concerning the number of reachable participants for each participant. 
\item We present first insights on the theoretical maximum spread of information possible within the communication network emerging from code review: the number of reachable participants. 
\item We highlight possible probabilistic extensions to and future applications of our model as a proxy for the capacity of code review as a knowledge-sharing platform.
\end{itemize}

The manuscript is structured as follows: We begin with a gentle mathematical introduction to time-varying hypergraphs in Section~\ref{sec_primer}. In Section~\ref{sec_background}, we provide an overview of state of the art on graph-based communication models in software engineering and related disciplines, as well as in-silico experiments and simulation in software engineering. We formalize code review as channel-based communication to the extent we deem necessary and define our conceptual and computer model in Section~\ref{sec_model}. In Section~\ref{sec_experimental_design}, we showcase and validate our model in a computer simulation rendering an artificial information diffusion in an empirical communication network that emerges from code review at Microsoft. After we present the resulting comparison of the time-ignoring and time-respecting reachable participants in an information diffusion simulation in Section~\ref{sec_results} and discuss the findings in Section~\ref{sec_discussion}, the manuscript closes with a conclusion and future work in Section~\ref{sec_conclusion}.

\section{A gentle introduction to time-varying hypergraphs}\label{sec_primer}

In this work, we combine two lesser-known graph-theoretical concepts: time-variance of graphs and hypergraphs. We follow the definitions and notation by \citet{Casteigts2012} for time-varying graphs  and by \citet{Ouvrard2020} for hypergraphs to a large extent. 

A \emph{time-varying graph} is a graph whose edges (and vertices) are active or available only at specific points in time. A \emph{hypergraph} is a generalization of a graph in which an edge (a so-called \emph{hyperedge}) can connect any arbitrary number\footnote{A classical graph is a subclass of a hypergraph with hyperedges that always connect exactly two (in case of self-loops not necessarily distinct) vertices.} of vertices. 

Thus, a time-varying hypergraph is a hypergraph which hyperedges (and vertices) are time-dependent. Mathematically, a time-varying hypergraph is a quintuple $\mathcal{H} = (V, \mathcal{E}, \rho, \xi, \psi)$ where 
\begin{itemize}
\item $V$ is a set of vertices,
\item $\mathcal{E}$ is a set of hyperedges connecting any number of vertices,
\item $\rho$ is the \emph{hyperedge presence function} indicating whether a hyperedge is active at a given time,
\item $\xi \colon E \times \mathcal{T} \rightarrow \mathbb{T}$ is the \emph{latency function} indicating the duration to cross a given hyperedge,
\item $\psi \colon V \times \mathcal{T} \rightarrow \{0, 1\}$ is the \emph{vertex presence function} indicating whether a given vertex is available at a given time, and 
\item $\mathcal{T} \in \mathbb{T}$ is the lifetime of the system.
\end{itemize}

The temporal domain $\mathbb{T}$ is generally assumed to be $\mathbb{N}$ for discrete-time systems or $\mathbb{R}_+$ for continuous-time systems.

Because the edges are time-dependent, the walk through a (hyper)graph is also time-dependent. Formally, a sequence of tuples 
\[\mathcal{J} = (e_1, t_1), (e_2, t_2), \dots, (e_k, t_k),\] 
such that $e_1, e_2, \dots, e_k$ is a walk in $\mathcal{H}$, is a \emph{journey} in $\mathcal{H}$ iff $\rho(e_i, t_i) = 1$ and $t_{i+1} > t_i + \xi(e_i, t_i)$ for all $i < k$.\footnote{We deviate from \citet{Casteigts2012} who require $t_{i+1} \geq t_i + \xi(e_i, t_i)$.} Additional constraints maybe required in specific domains of application, such as the condition $\rho_{[t_i, t_i+\xi(e_i,t_i)]}(e_i) = 1$: the hyperedge remains present until the hyperedge is crossed.

We define $\mathcal{J}^*_{\mathcal{H}}$ the set of all possible journeys in a time-varying graph $\mathcal{H}$ and $\mathcal{J}^*_{(u, v)} \in \mathcal{J}^*_{\mathcal{H}}$ the journeys between vertices $u$ and $v$. If $\mathcal{J}^*_{(u, v)} \neq \emptyset$, $u$ can reach $v$, or in short notation $u \leadsto v$. In general, journeys are not symmetric and transitive---regardless of whether the hypergraph is directed or undirected: $u \leadsto v \centernot\Leftrightarrow v \leadsto u$. Given a vertex $u$, the set $\{v \in V \colon u \leadsto v \}$ is called \emph{horizon} of vertex $u$.\footnote{The horizon of a vertex $v$ in a time-varying graph is not equivalent to the connected component that contains the vertex $v$ (neither strongly nor weakly component of the time-varying graph \cite{Nicosia2012}): the horizon is neither a reflexive (i.e.\ horizon of vertex $v$ does not necessarily contain the vertex $v$.) nor a symmetric relation.} 

%

A time-varying hypergraph $\mathcal{H}$ can be transformed in an equivalent bipartite graph $B = (V, \mathcal{E}, E, \psi)$ where 

\begin{itemize}
\item $V$ is the set of vertices from the equivalent hypergraph,
\item $\mathcal{E}$ is the set of hyperedges from the equivalent hypergraph, 
\item $V$ and $\mathcal{E}$ are disjunct ($V \cap \mathcal{E} = \emptyset$) and both vertices of the bipartite graph,
\item $E = \{(v, e) \;|\; u \in V, e \in \mathcal{E}\}$ are the vertices of the bipartite graph that connect vertices $V$ with hyperedges $\mathcal{E}$, and
\item $\psi$ is the edge presence function for the vertices $\mathcal{E}$ reflecting the edge presence function $\rho$ of the time-varying hypergraph such that $\psi(e) = \rho(e), e \in \mathcal{E}$.
\end{itemize}

Although an equivalent bipartite graph can represent a hypergraph, both concepts are semantically different. For an in-depth mathematical discussion, we refer the reader to the work by \citet{Ouvrard2020}. 

\begin{figure*} \centering
\begin{subfigure}[t]{\columnwidth} \centering
\input{figures/hypergraph.tex}
\caption{An example time-varying hypergraph, a generalization of a graph which edges, the so-called hyperedges, (denoted by $v_\square$) can link any arbitrary number of vertices (denoted by $e_\square$): For example, hyperedge $e_3$ connects four vertices. The reachability (or information diffusion in our case) of vertex $v_1$ depends highly on the temporal order of the hyperedges: if $e_1 < e_2 < e_4 < e_3$, the resulting horizon contains all vertices; if $e_1>e_2\geq e_3$ no information can be spread because no time-respecting path (journey) is available.}
\label{fig_example_hypergraph}
\end{subfigure} \hfill
%
%
\begin{subfigure}[t]{\columnwidth} \centering
\input{figures/bipartite_graph.tex}
\caption{Any hypergraph can be transformed into an equivalent bipartite graph: The hyperedges and the vertices from the time-varying hypergraph from Figure~\ref{fig_example_hypergraph} become the two distinct sets of vertices of a bipartite graph.}
\end{subfigure}

\caption{A simple example of a hypergraph and its bipartite-graph equivalent.}
\label{fig_intro}
\end{figure*}
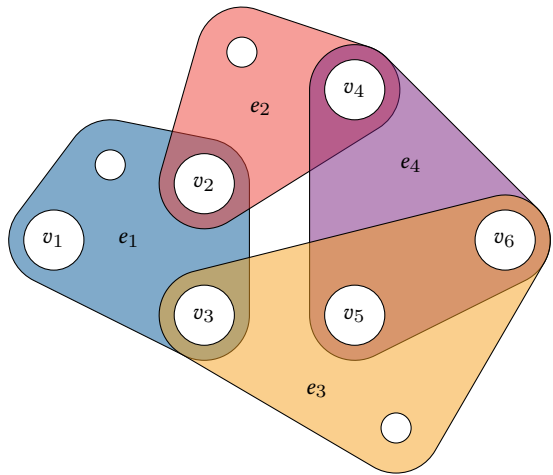
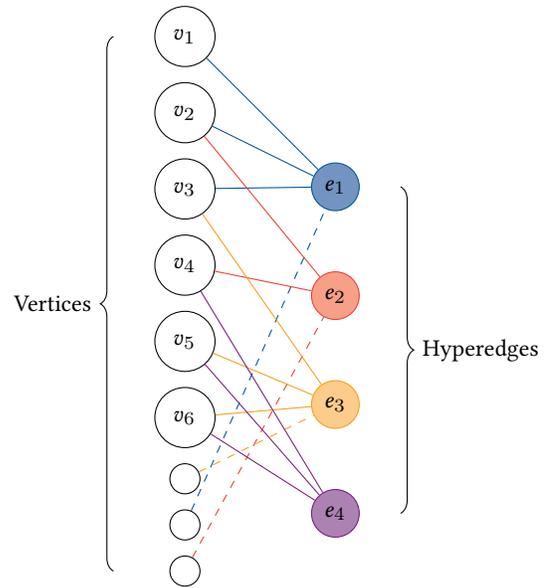

Figure~\ref{fig_intro} provides an example of a time-varying hypergraph and its transformation to an equivalent bipartite graph: The colors of the hyperedges reflect the colors of the righthand vertices in the bipartite graph. Furthermore, the example also shows the impact of time on the horizon of vertices in such hypergraphs. Depending on the presence of the hyperedges, there are different journeys from the vertices $v_1$ to $v_6$. Please mind that there is no time-respecting path (journey) in the opposite direction from $v_6$ to $v_1$. 

\section{Background}\label{sec_background} 

This section discusses related work on modelling communication in software engineering and provides context for our research approach, \emph{in-silico} experiments and simulations.

\subsection{Modelling communication in software engineering research}

To the best of our knowledge, our modelling approach for communication using time-dependent hypergraphs has not previously been used in software engineering research and other disciplines. Time-dependent hypergraphs are first applied in the research context of epidemiology: Independent of us and in parallel to our work, \citet{Antelmi2020} first defined and used time-depending hypergraphs to show the importance of time on disease diffusion. Neither the domain of epidemiology---information does not spread like viruses---nor the representation of hyperedges map to our research: In their model, hyperedges refer to geo-locations that are constant over time and vertices to persons meeting at those geo-locations over time. Hyperedges in our model reflect the channels of the information exchange and are time-dependent.

Software engineering research uses traditional graphs to model different types of information exchange and the networks that emerge from that communication. \citet{Herbold2021} identified in their systematic mapping study 182 studies researching various networks of developers modelled as graphs. We found that all studies use time-aggregated graphs to model developer interactions. Those limitations make time-aggregated graphs incapable of rendering time-dependent phenomena without introducing a large error by this simplification. 

However, the use of time-respecting network models and the research on information diffusion in software engineering is new but not wholly unexplored. 

\citet{Lamba2020} used a multi-layered time-dependent graph for investigating the tool diffusion of 12 quality assurance tools within the \emph{npm} ecosystem---without explicitly using this terminology. Although there are several similarities to our work at first sight, the used theoretical framework on the diffusion of innovations by \citet{Rogers2003} does not apply in the general case of communication as \citet{Rogers2003} states: Diffusion of innovations is ``a special type of communication, in that the messages are concerned with new ideas. Communication is a process in which participants create and share information to reach a mutual understanding.'' \cite{Rogers2003} Since we are modelling the exchange of information in general without any prior knowledge of its novelty value, the theory framework by Rogers does not apply to our research. Therefore, we explicitly use the term information diffusion in this study. 

\citet{Nia2010} investigated edge transitivity and the introduced error through the aggregation over time. They showed for the mailings lists of three open-source projects (Apache, Perl, and MySQL) that the clustering coefficient and the 2-path counts are robust to data aggregation across large intervals (over one year) even though such aggregation may lead to transitive faults. However, the results are only valid for time-aggregated systems. This implies that the findings do not apply to our research on and the modelling of information diffusion as the spread of information is a highly dynamic, time-dependent process. 

\citet{Gote2021} analyzed the temporal co-editing networks in software development teams using a rolling window approach \cite{Gote2021}. In detail, the study uses time-stamped bipartite graphs to model the relationship between developers and edited files. Since hypergraphs can be represented by bipartite graphs where hyperedges and vertices are the two distinct sets of vertices, the modelling approach is quite similar. However, the team converted this bipartite graph into a directed, acyclic graph (DAG) representing a sequence of consecutive co-editing relations of developers editing the given file to estimate knowledge flow. The nodes in this DAG represent commits and edges co-editing relationship between the authors of the commits. The connected components\footnote{We assume the model uses the \emph{strongly connected components} since the \emph{connected components} only exist in undirected graphs.} of the DAG represent proxies of knowledge flow, what we call information diffusion. Although this modelling approach respects the temporal order, the DAG cannot reflect the temporal distance and, thus, does not allow insights into how much time has left during the diffusion process. Only the topological distance (how many hops between two vertices) is available. The temporal distance (how much time has passed), however, reveals key characteristics of information since information ages constantly and information not delivered at the right time is outdated or simply invalid. The results always refer to the observation window but no more fine-grained insights. This shortcoming applies to all modelling approaches using any type of directed graph representing the order. 

A similar problem occurs with models based multigraphs for representing the parallel connection of multiple nodes. Although technically possible, multigraphs blur the relationship between an edge and a communication channel (i.e., code review): a communication channel would no longer correspond to one edge but a set of edges.

\subsection{In-silico experiments and simulations in software engineering research}

In our study, we conduct an \emph{in-silico} experiment. In contrast to in-vivo, in-vitro, and in-virtuo experiments, an in-silico experiment is performed solely via a computer simulation. Both subjects and the real world are described as simulation models \cite{Horta2003}. Any human interaction is reduced to a minimum. Those simulation models are like virtual laboratories where hypotheses about observed problems can be tested, and corrective policies can be experimented with before they are implemented in the real system \cite{Muller2008}. 

The use of the term \emph{simulation} varies substantially, from discipline and context \cite{DeFranca2020}. In this work, we rely on the definition by \citet{Banks2010}: A simulation is the imitation of the operation of a real-world process or system over time. The behavior of a system as it evolves over time is studied by developing a simulation model, a purposeful abstraction of a real-world system in the form of a set of assumptions concerning the system's operation. 

Although simulation models have been applied in different research fields of software engineering, e.g., process engineering, risk management, and quality assurance \cite{Muller2008}, due to the need for a large amount of knowledge, in-silico studies are scarce in software engineering \cite{Horta2003}.

\section{Modelling information diffusion with time-varying hypergraphs}\label{sec_model}

Communication is a complex and manifold process that changes over time. We need models as a purposeful and simplified abstraction of such complex phenomena, imitating those complex real-world processes to enable measurability, gain insights, predict outcomes, and understand the mechanics. 

A simulation model has two components: a conceptual model and a computer model \cite{DeFranca2016}. A conceptual model is a (non-software) abstraction of the simulation model that is to be developed, describing objectives, inputs, outputs, content, assumptions, and simplifications of the model \cite{Robinson2006}. On the other hand, the computer model describes the conceptual model implemented in software. 

In the following subsection, we define and discuss the conceptual and computer model of information diffusion in code review.

\subsection{Conceptual model}

Communication, the purposeful, intentional, and active exchange of information among humans, does not happen in the void. It requires a channel to exchange information. A \emph{communication channel} is a conduit for exchanging information among communication participants. Those channels are 

\begin{enumerate}
\item {\itshape multiplexing}---A channel connects all communication participants sending and receiving information.
\item {\itshape reciprocal}---The sender of information also receives information and the receiver also sends information. The information exchange converges. This can be in the form of feedback, queries, or acknowledgments. Pure broadcasting without any form of feedback does not satisfy our definition of communication.
\item {\itshape concurrent}---Although a human can only feed into and consume from one channel at a time, multiple concurrent channels are usually in use. 
\item {\itshape time-dependent}---Channels are not always available; the channels are closed after the information is transmitted.
\end{enumerate}

Channels group and structure the information for the communication participants over time and content. Over time, the set of all communication channels forms a communication network among the communication participants.

In the context of researching the information diffusion in this study, a communication channel is a discussion in a merge (or pull) request. A channel for a code review on a merge request begins with the initial submission and ends with the merge in case of an acceptance or a rejection. All participants of the review of the merge request feed information into the channel and, thereby, are connected through this channel and exchange information they communicate. After the code review is completed and the discussion has converged, the channel is closed and archived, and no new information becomes explicit and could emerge. However, a closed channel is usually not deleted but archived and is still available for passive information gathering. We do not intend to model this passive absorption of information from archived channels by retrospection with our model. 

From the previous postulates on channel-based communication in software engineering, we derive our computer model: Each communication medium forms an undirected, time-varying hypergraph in which hyperedges represent communication channels. Those hyperedges are available over time and make the hypergraph time-dependent. Additionally, we allow parallel hyperedges\footnote{This makes the hypergraph formally a \emph{multi-hypergraph} \cite{Ouvrard2020}. However, we consider the difference between a hypergraph and a multi-hypergraph as marginal since it is grounded in set theory: Sets do not allow multiple instances of the elements. Therefore, instead of a set of hyperedges, we use a multiset of hyperedges that allows multiple instances of the hyperedge.}---although unlikely, multiple parallel communication channels can emerge between the same participants at the same time but in different contexts.

Such an undirected, time-varying hypergraph reflects all four basic attributes of channel-based communication: 

\begin{itemize}
\item {\itshape multiplexing}---since a single hyperedge connects multiple vertices,
\item {\itshape concurrent}---since (multi-)hypergraphs allow parallel hyperedges,
\item {\itshape reciprocal}---since the hypergraph is undirected, information is exchanged in both directions, and
\item {\itshape time-dependent}---since the hypergraph is time-varying. 
\end{itemize}

In detail, we define our model for information diffusion in an observation window $\mathcal{T}$ to be an undirected time-varying hypergraph 
\[\mathcal{H} = (V, \mathcal{E}, \rho, \xi, \psi)\] 
where

\begin{itemize}
\item $V$ is the set of all human participants in the communication as vertices
\item $\mathcal{E}$ is a multiset (parallel edges are permitted) of all communication channels as hyperedges, 
\item $\rho$ is the \emph{hyperedge presence function} indicating whether a communication channel is active at a given time,
\item $\xi \colon E \times \mathcal{T} \rightarrow \mathbb{T}$, called \emph{latency function}, indicating the duration to exchange an information among communication participants within a communication channel (hyperedge),
\item $\psi \colon V \times \mathcal{T} \rightarrow \{0, 1\}$, called \emph{vertex presence function}, indicating whether a given vertex is available at a given time.
\end{itemize}

\begin{figure}\centering
\input{figures/observation_window.tex}
\caption{Not all communication channels started or ended within the observed time window (indicated by \textcolor{ACMDarkBlue}{blue}): Cut channels (indicated by \textcolor{ACMOrange}{orange}) are incomplete and lead to a blur at the borders of our measurements.}
\label{fig_observation_window}
\end{figure}
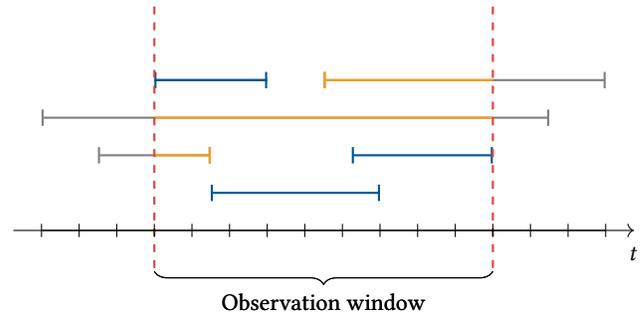

Communication and the spread of information are usually ongoing, continuous processes. As for any continuous, real-world process, we only can make assumptions about windowed observations of that phenomenon. The lifetime of our system is limited by this observation window which borders induce blur in our investigations: The communication may have started before or ended after our observed time window; information is lost. Thus, we must define our model's lifetime as the observation window. Figure~\ref{fig_observation_window} illustrates this problem of the observation window for an ongoing, continuous system. 

\subsection{Computer model}

We implement the hypergraph as an equivalent bipartite graph using the widely used Python graph package \emph{networkx} \cite{Hagberg2008}: The hypergraph vertices and hyperedges become two sets of vertices of the bipartite graph. The vertices of those disjoint sets are connected if a hypergraph edge was part of the hyperedge. For a more detailed and graphical description of the equivalence of hypergraphs and bipartite graphs, we refer the reader to  Section~\ref{sec_primer}.

To ensure that the computational model accurately represents the underlying mathematical model and its solution, we applied four quality assurance approaches in the model verification phase: 

\begin{itemize}
\item \textbf{Code walk-throughs}---We independently conducted code walk\hyp{}throughs through the simulation code with three Python and graph experts. 
\item \textbf{High test-coverage}---The simulation code has a test coverage of about $99 \%$. 
\item \textbf{Code readability and documentation}---We provide comprehensive documentation on the usage and design decisions to enable broad use and further development. We followed the standard Python programming style guidelines PEP8 for readability. 
\item \textbf{Publicly available and open source}---The model parameterization and simulation code \cite{Dorner2022software} as well as all intermediate and final results \cite{Dorner2022data} are publicly available.
\end{itemize}

\section{Experimental design}\label{sec_experimental_design}

In this section, we describe the simulation as an \emph{in-silico} experiment \cite{Felderer2020} that evaluates the impact of ignoring and respecting time in a temporal graph for modelling information diffusion in code review. The purpose of this simulation is two-fold: We provide a proof of concept of our modelling approach and present a first validation by comparison to another model \cite{DeFranca2020}. Through this comparison, the impact of time on communication networks becomes evident. 

In this computer simulation, we measure the number of individuals receiving information from a code review directly and indirectly in a best-case scenario. 

\begin{figure*} \centering
\input{figures/overview.tex}
\caption{An overview of the simulation. }
\label{fig_simulation_overview}
\end{figure*}
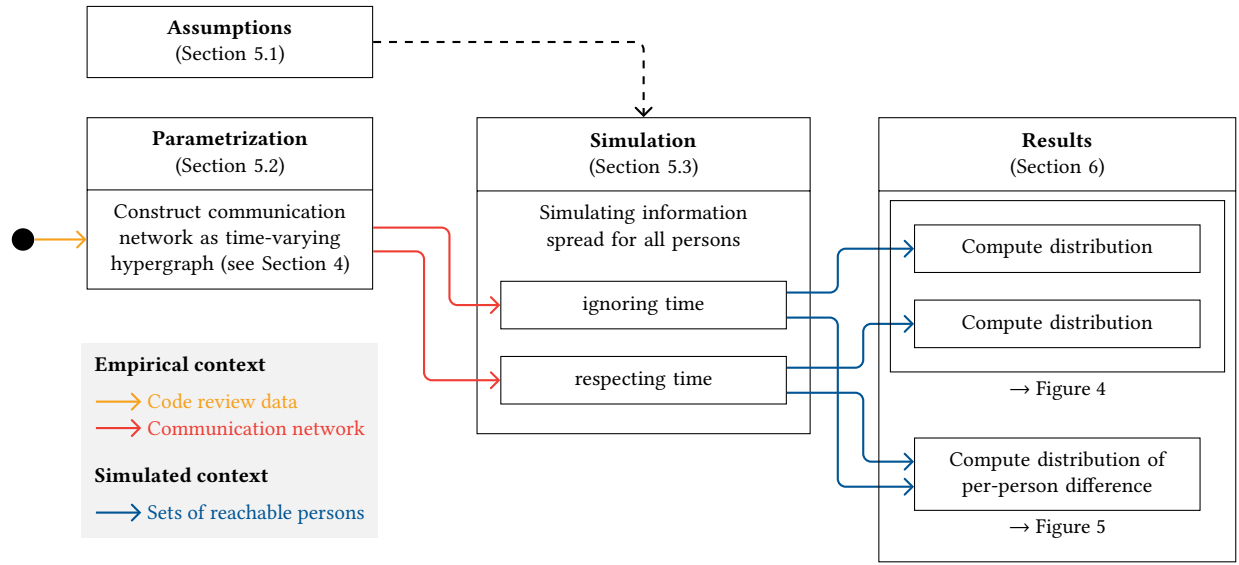

In Figure~\ref{fig_simulation_overview}, we present a high-level overview of our simulation. 

In the following subsection, we describe our simulation assumptions (Section~\ref{sec_assumptions}), the parametrization of our model using empirical data from code review at Microsoft (Section~\ref{sec_parametrization}), and the simulation mathematically and algorithmically (Section~\ref{sec_simulation}).

\subsection{Assumptions}\label{sec_assumptions}

For this study, we made the following assumptions for information diffusion in code review:

\begin{itemize}
\item {\itshape Channel-based}---Information can only be exchanged along the information channels. 
\item {\itshape Perfect caching}---All code review participants can remember and cache all information in all code reviews they participate in within the considered time frame. 
\item {\itshape Perfect diffusion}---All participants instantly pass on information at any occasion in all available communication channels in code review. 
\item {\itshape Information diffusion only in code review}---For this simulation, we assume that information gained from discussions in code review diffuses only through code review. 
\item {\itshape Information availability}---To have a common starting point and make the results comparable, the information to be diffused in the network is already available to the participant that is the origin of the information diffusion process. 
\end{itemize}

We discuss the impact of those assumptions in Section~\ref{sec_discussion}. The assumption $\psi \rightarrow 1$, meaning all code review participants are available over the considered time-frame $\mathcal{T}$ of four weeks, is implicit and does not impact the measurements: If a participant is either inactive or removed temporarily or permanently from the communication network has no impact on the number of reachable participants.

Our assumptions make the number of reachable participants a best-case scenario and do not likely represent an empirical information diffusion process. However, the relative comparison is adequate since all assumptions are equal for time-ignoring and time-respecting information diffusion measurements.

\subsection{Parametrization}\label{sec_parametrization}

To parametrize the model, we extracted all internal, human code review interactions tracked by Microsoft's internal code review tool \emph{CodeFlow} \cite{Bosu2015} run by Azure DevOps service. Although not Microsoft's only code review tool, it represents a large portion of the company's code review activity. All non-human code-review participants and interactions are excluded. The dataset contains all human code review interactions from 2020-02-03 to 2020-03-01, inclusively, -- corresponding to full four calendar weeks without significant discontinuities by public holidays such as Christmas. The time frame is arbitrary, however. 

The underlying hypergraph has $37,103$ vertices (developers) and $309,740$ hyperedges (communication channels) for both models. We made all code and data publicly available in our replication package.

\subsection{Simulation}\label{sec_simulation}

In our simulation, we use our parametrized communication model to measure how many participants can be reached from each participant using either time-respecting or time-ignoring paths in the communication network. 

Mathematically, the number of reachable participants is a set of vertices that can be reached from $u$:
\[\left| \left\{v \in V \colon u \leadsto v \right\} \right|\]  
If reachability is time-respecting, the measure is called \emph{horizon}. If time and the temporal order are ignored for the reachability, the horizon becomes the \emph{connected component} containing vertex $u$. 

Algorithmically, both measurements on the number of reachable participants for all vertices are variations of the breadth\hyp{}first search. Algorithm \ref{alg_horizon} describes the time-ignoring and time\hyp{}respecting breadth\hyp{}first search approach in pseudocode; our Python implementation can be found in the replication package. 

\begin{algorithm}
\DontPrintSemicolon
\SetKwInOut{Input}{Input}
\SetKwInOut{Output}{Output}
\Input{Time-varying Hypergraph $\mathcal{H} = (V, \mathcal{E}, \rho, \xi, \psi)$ \newline Start node $s \in V$}
\Output{An set for all node $v \in V$ reachable of $s$}
\BlankLine\; 
$Q \longleftarrow$ initialize empty queue\;
push $s \longrightarrow Q$\;
mark $s$ as reachable\;
\BlankLine\; 
\While{$Q \neq \emptyset$}{
	pop $Q \longrightarrow v$\;
	$N \longleftarrow \begin{cases}N(v) & \text{if time-ignoring}\\ \{n \in N(v) \subseteq V\;|\; v \leadsto n \} & \text{if time-respecting} \end{cases}$ \;
	\ForEach{$n \in N$\Comment*[f]{All available neighbors of $s$}}{
		\If{$n$ not marked as reachable}{
			push $n \longrightarrow Q$\;
			marks $n$ as reachable
		}
	}
}
\Return all reachable nodes \;
\caption{Breadth-first search for vertex $s$ of a time-varying hypergraph $\mathcal{H}$.}
\label{alg_horizon}
\end{algorithm}

The algorithm is integrated into our computer model and implemented in Python. All code is publicly available \cite{Dorner2022software}\footnote{For more information, see also \url{https://github.com/michaeldorner/only-time-will-tell}.} under MIT license. To ensure the correctness of the implementation, we created an extensive test setup.

\section{Results}\label{sec_results}

\begin{figure*} \centering
\input{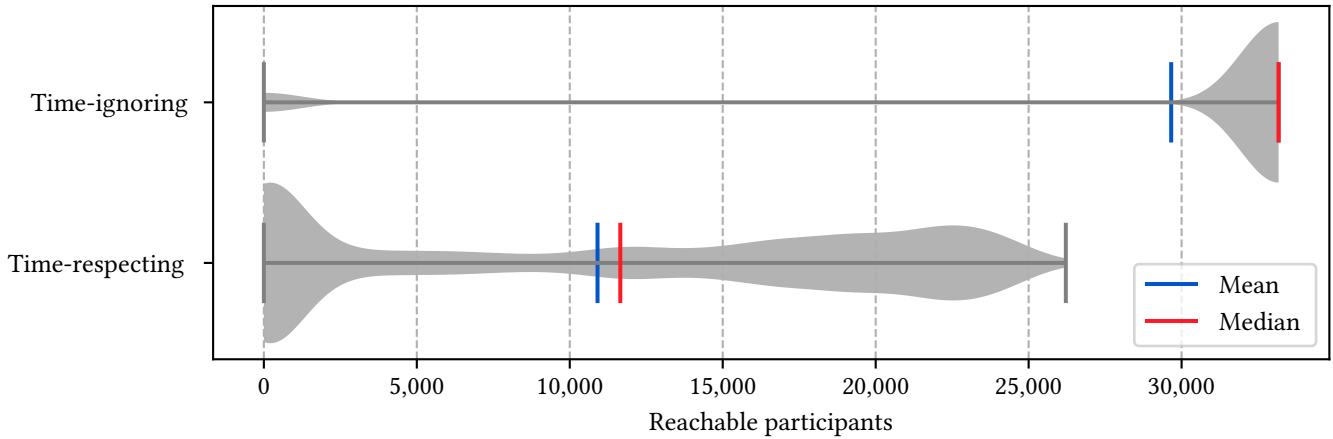}
\caption{Distribution of the time-respecting and time-ignoring reachable participants for a simulated information diffusion: All statistical locations (mean, median, min, and max) are significantly smaller when respecting time.}
\label{fig_dist}
\end{figure*}

All statistical locations of the reachable participants (namely median, mean, min, and max) are significantly smaller in the time-respecting information diffusion than the time-ignoring information diffusion: The mean time-ignoring reachable participants are $29,660$ persons, the median is $33,172$ persons. This unequal distribution is caused by the symmetry characteristic of the reachability in (undirected) graphs: All vertices have the same connected component. The largest component has $33,173$ persons, which is $89.41\%$ of all persons due to the symmetry characteristic of the connected component in (undirected) graphs. All other connected components are significantly smaller: The second-largest component has $108$ persons ($0.29\%$). The number of time-respecting reachable participants draws a more fine-grained picture: On average, $10,907$ persons (mean) and $11,652$ persons (median), respectively, can be reached. At most, $26,216$ persons ($70.66\%$) can be reached. Figure~\ref{fig_dist} contrasts both distributions.

\begin{figure*} \centering
\input{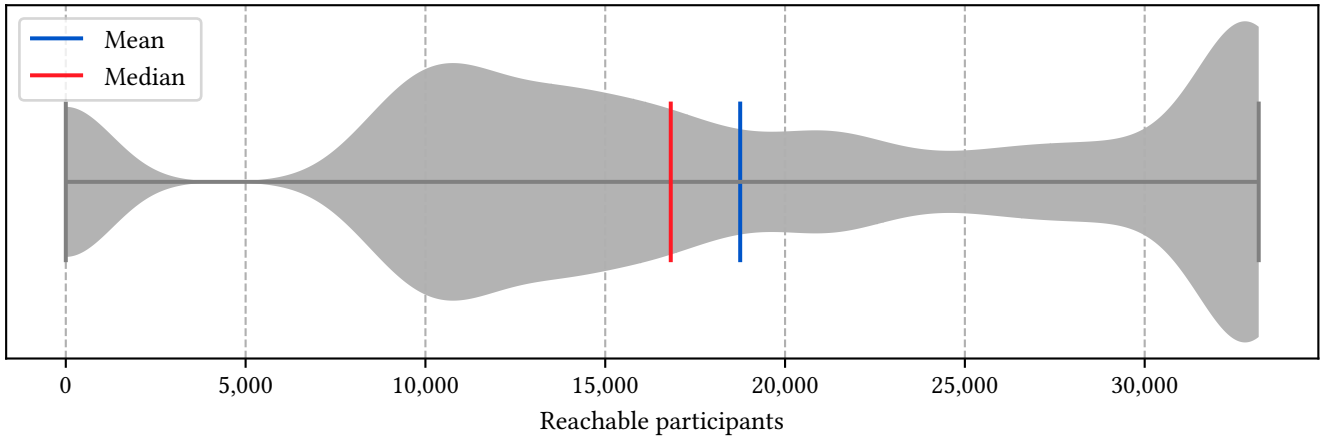}
\caption{Distribution of the per-person difference of the time-ignoring and time-respecting reachable participants: All persons have a significantly larger number of reachable persons if respecting time.}
\label{fig_diff}
\end{figure*}

The time-respecting and the time-ignoring per-person difference in the reachable participants differ significantly. In average, the difference is $18,752$ persons (mean) or $16,822$ persons (median). The largest per-person difference is $33,171$, which is also the maximum value: while all persons are reachable when time is ignored, no person is reachable when time is considered, i.e., no path in temporal order (journey) is available. Figure~\ref{fig_diff} depicts the per-person difference between the time-ignoring and time-respecting reachable participants. 

Both perspectives on reachable participants confirm the remarkable difference in respecting or ignoring time for measurements of information diffusion: Time-ignorance overestimates the available and temporal valid paths (journeys) in communication networks. The temporal order has a significant impact on the horizon and, thus, on the paths valid for information to diffuse.

\section{Discussion}\label{sec_discussion}

At this point, we would like to emphasize again that the measurements do not describe an empirical information diffusion: Although the network structure is constructed by real-world data and, therefore, empirical, the resulting information diffusion, the spread of information, is simulated. However, although the spread of information is artificial and the information has never empirically reached the participants, we believe the maximum number of reachable participants can be considered empirical, not neglecting the constraints we put on our simulation in the forms of our assumptions. 

All five assumptions described in Section~\ref{sec_experimental_design} are applied to both models. The constraints apply to both measurements to the same extent. Therefore, comparing the results ignoring and respecting time is sound and adequate. 

Our assumptions are not easily transferrable to other empirical investigations on information diffusion in code review or general. Both simulation assumptions of perfect caching and perfect diffusion are best-case assumptions leading to an upper bound. We strongly believe that this upper bound of reachable participants is not achievable and less meaningful in reality, particularly over larger time frames: information may get outdated, irrelevant, or even false over time. Also, human retentiveness, attention, and memory are limited. Future research can investigate the average number of reachable participants within code review and the impact of the topological and temporal distance on the probability of information diffusion. 

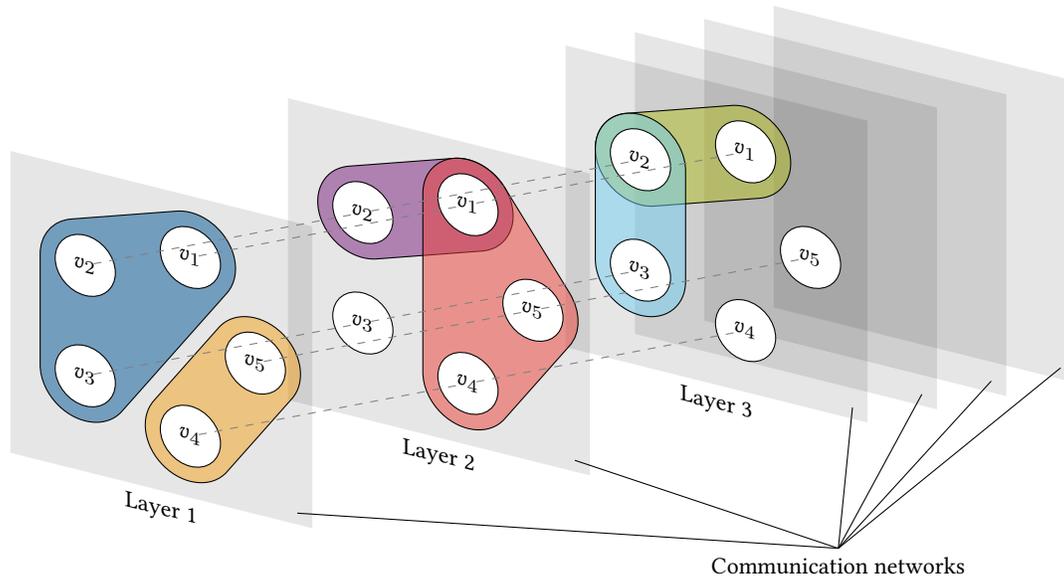
\begin{figure*}\centering
\input{figures/holistic_hypergraphs.tex}
\caption{For a more holistic view of information diffusion, different communication network layers are required.}
\label{fig_multilayer_networks}
\end{figure*}

Furthermore, information is not only diffused through code review but also through other communication media like instant messaging or virtual or in-person meetings. For a more holistic view of information diffusion in software engineering projects---not only through and within code review---, we need to capture more communication networks (e.g., code review, instant messaging, e-mail, classical meetings) stacked on each other to capture all possible information diffusion journeys. 
Figure~\ref{fig_multilayer_networks} gives an example of stacked communication networks consisting of overlapping hypergraphs. 

The second advantage of our model to be capable of rendering interconnections between more than two persons is not further discussed yet. 33.98\% of all code reviews at Microsoft involves more than two persons and, thus, cannot be captured by classical graphs, only by multigraphs having parallel edges. However, models with multigraphs---although technically possible---blur the relationship between an edge and a communication channel (i.e., code review): a code review would no longer correspond to one (hyper)edge but a set of edges. The graph becomes less expressive and more complex to compute and simulate.

\section{Conclusion}\label{sec_conclusion}

We present a model based on time-varying hypergraphs for modelling and analyzing information diffusion within code review. The model overcomes the limitations of existing graph-based models and enables research on time-respecting and multilateral information diffusion. 

Our simulation based on the code review at Microsoft to estimate the empirical impact of time-dependency on information diffusion reveals that significantly fewer code review participants are reachable and, therefore, significantly fewer paths to diffuse information are valid if time is respected. Ignoring time in communication networks introduces a large error since the time-ignoring model overestimates the available and valid paths within such communication networks. 

We believe that the available information diffusion paths, as well as the topological (measured in the number of time-respecting hops) or temporal distance (measured in time) between participants revealed by our model, provide a solid foundation for future research on the capacity of code review as a knowledge-sharing platform as suggested by prior qualitative studies \cite{Bacchelli2013, Bosu2017, Baum20161}.

Our model can be easily extended by probability, an integral part of information diffusion: not every information is spread on every occasion: \emph{random or probabilistic time-varying graphs} with an edge presence function $\rho \colon E \times \mathcal{T} \rightarrow [0,1]$ or vertex presence function $\psi \colon V \times \mathcal{T} \rightarrow [0, 1]$ allows to render probabilistic processes of information diffusion and estimate the stability of communication networks. As a generalization of a traditional graph, hypergraphs are a promising modelling tool for not only communication networks but also other higher-order systems since they are compatible with traditional graph metrics and algorithms.

To enable researchers and practitioners to replicate, reproduce, and extend our work and model, we provide an extensive replication package containing all code \cite{Dorner2022software} and data \cite{Dorner2022data}\footnote{For more information, see also \url{https://github.com/michaeldorner/only-time-will-tell}.}. We also explicitly encourage researchers from outside software engineering to apply, revise, and advance our model.

\begin{acks}
We thank Dietmar Pfahl for his valuable contributions to the presentation of the paper and Andreas Bauer and Ehsan Zabardast for their feedback which improved the simulation code greatly. We also thank the anonymous reviewers for their careful reading of our manuscript and their many insightful comments and helpful suggestions. 

This work was supported by the KKS Foundation through the SERT project (research profile grant 2018/010) at Blekinge Institute of Technology.
\end{acks}

\bibliographystyle{ACM-Reference-Format}
\bibliography{bibliography.bib}

\end{document}

%% file: tikz-hypergraph.tex
\def\rotateclockwise#1{
  \newdimen\xrw
  \pgfextractx{\xrw}{#1}
  \newdimen\yrw
  \pgfextracty{\yrw}{#1}
  \pgfpoint{\yrw}{-\xrw}
}

\def\rotatecounterclockwise#1{
  \newdimen\xrcw
  \pgfextractx{\xrcw}{#1}
  \newdimen\yrcw
  \pgfextracty{\yrcw}{#1}
  \pgfpoint{-\yrcw}{\xrcw}
}

\def\outsidespacerpgfclockwise#1#2#3{
  \pgfpointscale{#3}{
    \rotateclockwise{
      \pgfpointnormalised{
        \pgfpointdiff{#1}{#2}}}}
}

\def\outsidespacerpgfcounterclockwise#1#2#3{
  \pgfpointscale{#3}{
    \rotatecounterclockwise{
      \pgfpointnormalised{
        \pgfpointdiff{#1}{#2}}}}
}

\def\outsidepgfclockwise#1#2#3{
  \pgfpointadd{#2}{\outsidespacerpgfclockwise{#1}{#2}{#3}}
}

\def\outsidepgfcounterclockwise#1#2#3{
  \pgfpointadd{#2}{\outsidespacerpgfcounterclockwise{#1}{#2}{#3}}
}

\def\outside#1#2#3{
  ($ (#2) ! #3 ! -90 : (#1) $)
}

\def\cornerpgf#1#2#3#4{
  \pgfextra{
    \pgfmathanglebetweenpoints{#2}{\outsidepgfcounterclockwise{#1}{#2}{#4}}
    \let\anglea\pgfmathresult
    \let\startangle\pgfmathresult

    \pgfmathanglebetweenpoints{#2}{\outsidepgfclockwise{#3}{#2}{#4}}
    \pgfmathparse{\pgfmathresult - \anglea}
    \pgfmathroundto{\pgfmathresult}
    \let\arcangle\pgfmathresult
    \ifthenelse{180=\arcangle \or 180<\arcangle}{
      \pgfmathparse{-360 + \arcangle}}{
      \pgfmathparse{\arcangle}}
    \let\deltaangle\pgfmathresult

    \newdimen\x
    \pgfextractx{\x}{\outsidepgfcounterclockwise{#1}{#2}{#4}}
    \newdimen\y
    \pgfextracty{\y}{\outsidepgfcounterclockwise{#1}{#2}{#4}}
  }
  -- (\x,\y) arc [start angle=\startangle, delta angle=\deltaangle, radius=#4]
}

\def\corner#1#2#3#4{
  \cornerpgf{\pgfpointanchor{#1}{center}}{\pgfpointanchor{#2}{center}}{\pgfpointanchor{#3}{center}}{#4}
}

\def\hedgem#1#2#3#4{
  
  \outside{#1}{#2}{#4}
  \pgfextra{
    \def\hgnodea{#1}
    \def\hgnodeb{#2}
  }
  foreach \c in {#3} {
    \corner{\hgnodea}{\hgnodeb}{\c}{#4}
    \pgfextra{
      \global\let\hgnodea\hgnodeb
      \global\let\hgnodeb\c
    }
  }
  \corner{\hgnodea}{\hgnodeb}{#1}{#4}
  \corner{\hgnodeb}{#1}{#2}{#4}
  -- cycle
}

\def\hedgeii#1#2#3{
  \hedgem{#1}{#2}{}{#3}
}

%% file: figures/hypergraph.tex
\begin{tikzpicture}
\tikzstyle{important}=[draw, circle, minimum size=8mm, fill=white];
\tikzstyle{unimportant}=[draw, circle, minimum size=4mm, fill=white];

\node at (0,1) (v1) {};
\node at (2,1.75) (v2) {};
\node at (2,0) (v3) {};
\node at (4,3) (v4) {};
\node at (4,0) (v5) {};
\node at (6,1) (v6) {};
\node at (4.55,-1.5) (v7) {};
\node at (0.75,2) (v8) {};
\node at (2.5,3.5) (v9) {};

\draw[draw, fill=ACMDarkBlue, fill opacity=0.5] \hedgem{v1}{v8}{v2,v3}{6mm};
\draw[draw, fill=ACMRed, fill opacity=0.5] \hedgem{v2}{v9}{v4}{6mm};
\draw[draw, fill=ACMPurple, fill opacity=0.5] \hedgem{v5}{v4}{v6}{6mm};
\draw[draw, fill=ACMOrange, fill opacity=0.5] \hedgem{v3}{v6}{v7}{6mm};

\node[important] at (v1) {$v_1$};
\node[important] at (v2) {$v_2$};
\node[important] at (v3) {$v_3$};
\node[important] at (v4) {$v_4$};
\node[important] at (v5) {$v_5$};
\node[important] at (v6) {$v_6$};
\node[unimportant] at (v7) {};
\node[unimportant] at (v8) {};
\node[unimportant] at (v9) {};

\node at (1, 1) {$e_1$};
\node at (2.75, 2.75) {$e_2$};
\node at (3.5, -1) {$e_3$};
\node at (4.75, 2) {$e_4$};

\end{tikzpicture} 

%% file: figures/bipartite_graph.tex
\begin{tikzpicture}
	
\tikzstyle{important}=[draw, circle, minimum size=8mm, fill=white];
\tikzstyle{unimportant}=[draw, circle, minimum size=4mm, fill=white];

\begin{scope}[start chain=going below, node distance=2mm]
\foreach \i in {1,2,...,6}
  \node[on chain, important] (v\i) {$v_\i$}; 
\foreach \i in {7, 8, 9}
  \node[on chain, unimportant] (v\i) {};
\end{scope}

\begin{scope}[xshift=2cm, yshift=-2cm, start chain=going below, node distance=8mm]
\foreach \e/\c in {1/ACMDarkBlue, 2/ACMRed, 3/ACMOrange, 4/ACMPurple}
  \node[on chain, draw, circle, \c, fill=\c!50] (e\e) {\textcolor{black}{$e_\e$}};
\end{scope}

\draw[ACMDarkBlue] (v1) -- (e1);
\draw[ACMDarkBlue] (v2) -- (e1);
\draw[ACMDarkBlue] (v3) -- (e1);
\draw[ACMDarkBlue, dashed] (v8) -- (e1);

\draw[ACMRed] (v2) -- (e2);
\draw[ACMRed] (v4) -- (e2);
\draw[ACMRed, dashed] (v9) -- (e2);

\draw[ACMOrange] (v3) -- (e3);
\draw[ACMOrange] (v5) -- (e3);
\draw[ACMOrange, dashed] (v7) -- (e3);
\draw[ACMOrange] (v6) -- (e3);

\draw[ACMPurple] (v4) -- (e4);
\draw[ACMPurple] (v5) -- (e4);
\draw[ACMPurple] (v6) -- (e4);

\draw [decorate,decoration={brace, amplitude=5pt, mirror, raise=4ex}] (v1.west) -- ([xshift=-2mm]v9.west) node[midway, anchor=east, label={[xshift=-5mm]left:Vertices}]{};

\draw [decorate,decoration={brace, amplitude=5pt, raise=4ex}] (e1.east) -- (e4.east) node[midway, anchor=west, label={[xshift=5mm]right:Hyperedges}]{};

\end{tikzpicture}

%% file: figures/observation_window.tex
\tikzstyle{cr}=[draw, |-|, thick, gray]
\begin{tikzpicture}[x=0.75cm, y=1cm]
	
	\draw[cr, ACMDarkBlue] (4, 1/2) -- (7, 1/2); 
	
	\draw[cr] (2,2/2) -- (4, 2/2); 
	\draw[draw, thick, -|, ACMOrange] (3, 2/2) -- (4, 2/2);

	\draw[cr, ACMDarkBlue] (6.5, 2/2) -- (9, 2/2); 
	
	\draw[cr] (1,3/2) -- (10, 3/2); 
	\draw[draw, thick, ACMOrange] (3, 3/2) -- (9, 3/2); 

	\draw[cr] (6,4/2) -- (11,4/2); 
	\draw[draw, thick, |-, ACMOrange] (6, 4/2) -- (9, 4/2); 
	
	\draw[cr, ACMDarkBlue] (3,4/2) -- (5,4/2);
	\draw[draw, thick, ACMDarkBlue] (3,4/2) -- (5,4/2) ; 

	\draw[dashed, ACMRed, thick] (3,-0.5) -- (3,3);
	\draw[dashed, ACMRed, thick] (9,-0.5) -- (9,3);
	\draw [decorate,decoration={brace, amplitude=5pt, mirror, raise=4ex}] (3,0) -- (9,0) node[midway, yshift=-3em]{Observation window};
	\draw [decorate,decoration={brace, amplitude=5pt, mirror, raise=4ex}] (3,0) -- (9,0) node[midway, yshift=-3em]{Observation window};
	\draw[->] (0.5,0) -- (11.5,0) node[label=below:$t$] {};
	\draw [
    postaction={
        draw,
        decoration=ticks,
        segment length=0.5cm,
        decorate,
    }
 ] (1,0) -- (11,0);
\end{tikzpicture}

%% file: figures/overview.tex
\tikzset{%
	a/.style={
		-angle 90, 
		thick,
		rounded corners=3,
	},
	box/.style={
		rectangle,
		draw=black,
		minimum width=12em, 
		text width=10em,
		minimum height=2em,
		inner sep=5pt,
		text centered,
	},
}
\begin{tikzpicture}[font=\small, xscale=5.5]

\node[box] (assumptions) at (0, 2) {\textbf{Assumptions}\\(Section~\ref{sec_assumptions})};
\node[box] (ignore_time) at (1, -1.5) {ignoring time};
\node[box] (respect_time) at (1, -2.5){respecting time};

\node[box, rectangle split, anchor=north, rectangle split parts=2] (parametrization) at (0,1) { 
\nodepart{one}\textbf{Parametrization}\\ (Section~\ref{sec_parametrization})%
\nodepart{two} Construct communication network as time-varying hypergraph (see Section~\ref{sec_model})
};

\coordinate (c) at (-0.5, 1);
\node[circle, fill=black] (start) at (c|-parametrization.two west) {};

\node[box, minimum width=14em, anchor=north, rectangle split, rectangle split parts=2] (procedure) at (1, 1){
\nodepart{one}\textbf{Simulation}\\(Section~\ref{sec_simulation})
\nodepart{two} Simulating information spread for all persons\\[7em]$ $}; 

\node[box] (comp_dist_1) at (2, -0.75) {Compute distribution};
\node[box] (comp_dist_2) at (2, -1.75) {Compute distribution};

\node[box, label={[name=label_diff, text centered, label distance=0em]below:$\rightarrow$ Figure~\ref{fig_diff}}] (comp_diff) at (2, -3.75) {Compute distribution of per-person difference};

\draw[a, dashed] (assumptions.east) -| (procedure.north);
\draw[a, ACMOrange] (start) -- (parametrization.two west);
\draw[a, ACMRed] ([yshift=-0.5em]parametrization.two east) -| ($(parametrization.-5)!0.4!(respect_time.185)$) |- (respect_time.west);
\draw[a, ACMRed] ([yshift=0.5em]parametrization.two east) -| ($(parametrization.3)!0.6!(ignore_time.175)$) |- (ignore_time.west);

\draw[a, ACMDarkBlue] (ignore_time.5) -| ($(ignore_time.5)!0.4!(comp_dist_1.west)$) |- (comp_dist_1.west);
\draw[a, ACMDarkBlue] (respect_time.5) -| ($(respect_time.5)!0.6!(comp_dist_2.west)$) |- (comp_dist_2.west);

\draw[a, ACMDarkBlue] (ignore_time.-5) -| ($(ignore_time.5)!0.4!(comp_diff.185)$) |- (comp_diff.185);
\draw[a, ACMDarkBlue] (respect_time.-5) -| ($(respect_time.-5)!0.6!(comp_diff.175)$) |- (comp_diff.175);

\node[box, inner sep=1em, fit={(comp_dist_1.north east) (comp_dist_2.south west)}, label={[text centered, label distance=0em]below:$\rightarrow$ Figure~\ref{fig_dist}}] (fig4box) {};

\node[box, minimum width=15em, anchor=north, rectangle split, rectangle split parts=2] (results) at (2, 1){
\nodepart{one}\textbf{Results}\\(Section~\ref{sec_results})
\nodepart{two} \\[15em]$ $};

\node[draw=none, anchor=north, align=left, inner sep=5pt, fill=gray!10] (8888) at (0, -2){
\textbf{Empirical context}\\[0.5em]
\tikz[baseline=-0.75ex]{\draw[-angle 90, thick, ACMOrange] (0,0) -- (5ex,0); } \textcolor{ACMOrange}{Code review data} \\
\tikz[baseline=-0.75ex]{\draw[-angle 90, thick, ACMRed] (0,0) -- (5ex,0); } \textcolor{ACMRed}{Communication network} \\[1em]
\textbf{Simulated context}\\[0.5em]
\tikz[baseline=-0.75ex]{\draw[-angle 90, thick, ACMDarkBlue] (0,0) -- (5ex,0); } \textcolor{ACMDarkBlue}{Sets of reachable persons}
}; 
\end{tikzpicture}

%% file: figures/holistic_hypergraphs.tex
\tikzset{corners/.style={draw,fit={#1},rectangle,inner sep=0}}

\begin{tikzpicture}

\newcommand{\xslant}{0}
\newcommand{\yslant}{-0.25}

\newcommand{\xshift}{105}
\newcommand{\yshift}{20}

\tikzstyle{node}=[draw, circle, minimum size=8mm, fill=white];

\begin{scope}[
	yshift=2.75*\yshift, xshift=2.75*\xshift, 
	every node/.append style={yslant=\yslant, xslant=\xslant}, 
	yslant=\yslant, xslant=\xslant]

\node[draw=none, corners={(-2,-2.5) (2,2.5)}, fill=black, opacity=0.1] (layern) {};

\foreach \i in {1,...,5}{
	\node (vn\i) at (360/5 * \i:1.25cm) {};
}
\end{scope}

\begin{scope}[
	yshift=2.5*\yshift, xshift=2.5*\xshift, 
	every node/.append style={yslant=\yslant, xslant=\xslant}, 
	yslant=\yslant, xslant=\xslant]

\node[draw=none, corners={(-2,-2.5) (2,2.5)}, fill=black, opacity=0.1] (layer5) {};

\end{scope}

\begin{scope}[
	yshift=2.25*\yshift, xshift=2.25*\xshift, 
	every node/.append style={yslant=\yslant, xslant=\xslant}, 
	yslant=\yslant, xslant=\xslant]

\node[draw=none, corners={(-2,-2.5) (2,2.5)}, fill=black, opacity=0.1] (layer4) {};

\end{scope}

\begin{scope}[
	yshift=2*\yshift, xshift=2*\xshift, 
	every node/.append style={yslant=\yslant, xslant=\xslant}, 
	yslant=\yslant, xslant=\xslant]
\foreach \i in {1,...,5}{
	\node (v3\i) at (360/5 * \i:1.25cm) {};
}
\node[draw=none, corners={(-2,-2.5) (2,2.5)}, fill=black, opacity=0.1, label=below:Layer 3] (layer3) {};

\begin{scope}[fill opacity=0.5]
	\draw[draw, fill=ACMGreen] \hedgeii{v31}{v32}{6mm};
	\draw[draw, fill=ACMLightBlue] \hedgeii{v32}{v33}{6mm};
\end{scope}
\foreach \i in {1,...,5}{
	\node[node] at (360/5 * \i:1.25cm) {$v_\i$};
}
\node (p3) at (-3, -2) {};

\end{scope}

\begin{scope}[
	yshift=\yshift, xshift=\xshift, 
	every node/.append style={yslant=\yslant, xslant=\xslant}, 
	yslant=\yslant, xslant=\xslant]
\foreach \i in {1,...,5}{
	\node (v2\i) at (360/5 * \i:1.25cm) {};
}

\node[draw=none, corners={(-2,-2.5) (2,2.5)}, fill=black, opacity=0.1, label=below:Layer 2] (layer2) {};

\begin{scope}[fill opacity=0.5]
	\draw[draw, fill=ACMPurple] \hedgeii{v21}{v22}{6mm};
	\draw[draw, fill=ACMRed] \hedgem{v21}{v25}{v24}{6mm};
\end{scope}
\foreach \i in {1,...,5}{
	\node[node] at (360/5 * \i:1.25cm) {$v_\i$};
}
\node (p2) at (-3, -2) {};
\end{scope}

\begin{scope}[
	yshift=0, xshift=0, 
	every node/.append style={yslant=\yslant, xslant=\xslant}, 
	yslant=\yslant, xslant=\xslant]
\foreach \i in {1,...,5}{
	\node (v1\i) at (360/5 * \i:1.25cm) {};
}


\node[draw=none, corners={(-2,-2.5) (2,2.5)}, fill=black, opacity=0.1, label=below:Layer 1] (layer1) {};

\begin{scope}[fill opacity=0.5]
	\draw[draw, fill=ACMDarkBlue] \hedgem{v11}{v13}{v12}{6mm};
	\draw[draw, fill=ACMOrange] \hedgeii{v14}{v15}{6mm};
\end{scope}
\foreach \i in {1,...,5}{
	\node[node] at (360/5 * \i:1.25cm) {$v_\i$};
}
\end{scope}

\foreach \i in {1,...,5}{
	\draw[dashed, gray] (v1\i) -- (v2\i) --(v3\i); 
}

\node (communication_network) at (9, -3) {Communication networks};

\foreach \i in {1,..., 5, n}{
	\draw ([xshift=-2mm, yshift=2mm]layer\i.south east) -- (communication_network.north); 
}

%
%
%

\end{tikzpicture}

%% file: main.bbl

\begin{thebibliography}{23}


\ifx \showCODEN    \undefined \def \showCODEN     #1{\unskip}     \fi
\ifx \showDOI      \undefined \def \showDOI       #1{#1}\fi
\ifx \showISBNx    \undefined \def \showISBNx     #1{\unskip}     \fi
\ifx \showISBNxiii \undefined \def \showISBNxiii  #1{\unskip}     \fi
\ifx \showISSN     \undefined \def \showISSN      #1{\unskip}     \fi
\ifx \showLCCN     \undefined \def \showLCCN      #1{\unskip}     \fi
\ifx \shownote     \undefined \def \shownote      #1{#1}          \fi
\ifx \showarticletitle \undefined \def \showarticletitle #1{#1}   \fi
\ifx \showURL      \undefined \def \showURL       {\relax}        \fi
\providecommand\bibfield[2]{#2}
\providecommand\bibinfo[2]{#2}
\providecommand\natexlab[1]{#1}
\providecommand\showeprint[2][]{arXiv:#2}

\bibitem[\protect\citeauthoryear{Antelmi, Cordasco, Spagnuolo, and
  Scarano}{Antelmi et~al\mbox{.}}{2020}]%
        {Antelmi2020}
\bibfield{author}{\bibinfo{person}{Alessia Antelmi}, \bibinfo{person}{Gennaro
  Cordasco}, \bibinfo{person}{Carmine Spagnuolo}, {and}
  \bibinfo{person}{Vittorio Scarano}.} \bibinfo{year}{2020}\natexlab{}.
\newblock \showarticletitle{A design-methodology for epidemic dynamics via
  time-varying hypergraphs}.
\newblock \bibinfo{journal}{\emph{Proceedings of the International Joint
  Conference on Autonomous Agents and Multiagent Systems, AAMAS}}
  \bibinfo{volume}{2020-May} (\bibinfo{year}{2020}), \bibinfo{pages}{61--69}.
\newblock
\showISBNx{9781450375184}
\showISSN{15582914}
\urldef\tempurl%
\url{https://doi.org/10.5555/3398761.3398774}
\showDOI{\tempurl}


\bibitem[\protect\citeauthoryear{Bacchelli and Bird}{Bacchelli and
  Bird}{2013}]%
        {Bacchelli2013}
\bibfield{author}{\bibinfo{person}{Alberto Bacchelli} {and}
  \bibinfo{person}{Christian Bird}.} \bibinfo{year}{2013}\natexlab{}.
\newblock \showarticletitle{Expectations, outcomes, and challenges of modern
  code review}.
\newblock \bibinfo{journal}{\emph{Proceedings - International Conference on
  Software Engineering}} (\bibinfo{year}{2013}), \bibinfo{pages}{712--721}.
\newblock
\showISBNx{9781467330763}
\showISSN{02705257}
\urldef\tempurl%
\url{https://doi.org/10.1109/ICSE.2013.6606617}
\showDOI{\tempurl}


\bibitem[\protect\citeauthoryear{Banks, Carson, Nelson, and Nicol}{Banks
  et~al\mbox{.}}{2010}]%
        {Banks2010}
\bibfield{author}{\bibinfo{person}{Jerry Banks}, \bibinfo{person}{J.S. Carson},
  \bibinfo{person}{Barry~L Nelson}, {and} \bibinfo{person}{David~M Nicol}.}
  \bibinfo{year}{2010}\natexlab{}.
\newblock \bibinfo{booktitle}{\emph{Discrete event system simulation Solutions
  Manual} (\bibinfo{edition}{5} ed.)}.
\newblock \bibinfo{publisher}{Pearson Education}. 639 pages.
\newblock
\showISBNx{978-1-292-02437-0}


\bibitem[\protect\citeauthoryear{Baum, Liskin, Niklas, and Schneider}{Baum
  et~al\mbox{.}}{2016}]%
        {Baum20161}
\bibfield{author}{\bibinfo{person}{Tobias Baum}, \bibinfo{person}{Olga Liskin},
  \bibinfo{person}{Kai Niklas}, {and} \bibinfo{person}{Kurt Schneider}.}
  \bibinfo{year}{2016}\natexlab{}.
\newblock \showarticletitle{Factors influencing code review processes in
  industry}.
\newblock \bibinfo{journal}{\emph{Proceedings of the 2016 24th ACM SIGSOFT
  International Symposium on Foundations of Software Engineering - FSE 2016}}.
\newblock
\showISBNx{9781450342186}
\urldef\tempurl%
\url{https://doi.org/10.1145/2950290.2950323}
\showDOI{\tempurl}


\bibitem[\protect\citeauthoryear{Bosu, Carver, Bird, Orbeck, and Chockley}{Bosu
  et~al\mbox{.}}{2017}]%
        {Bosu2017}
\bibfield{author}{\bibinfo{person}{Amiangshu Bosu}, \bibinfo{person}{Jeffrey~C.
  Carver}, \bibinfo{person}{Christian Bird}, \bibinfo{person}{Jonathan Orbeck},
  {and} \bibinfo{person}{Christopher Chockley}.}
  \bibinfo{year}{2017}\natexlab{}.
\newblock \showarticletitle{Process Aspects and Social Dynamics of Contemporary
  Code Review: Insights from Open Source Development and Industrial Practice at
  Microsoft}.
\newblock \bibinfo{journal}{\emph{IEEE Transactions on Software Engineering}}
  \bibinfo{volume}{43} (\bibinfo{year}{2017}), \bibinfo{pages}{56--75}.
\newblock
Issue 1.
\showISSN{00985589}
\urldef\tempurl%
\url{https://doi.org/10.1109/TSE.2016.2576451}
\showDOI{\tempurl}


\bibitem[\protect\citeauthoryear{Bosu, Greiler, and Bird}{Bosu
  et~al\mbox{.}}{2015}]%
        {Bosu2015}
\bibfield{author}{\bibinfo{person}{Amiangshu Bosu}, \bibinfo{person}{Michaela
  Greiler}, {and} \bibinfo{person}{Christian Bird}.}
  \bibinfo{year}{2015}\natexlab{}.
\newblock \showarticletitle{Characteristics of useful code reviews: An
  empirical study at Microsoft}. In \bibinfo{booktitle}{\emph{IEEE
  International Working Conference on Mining Software Repositories}},
  Vol.~\bibinfo{volume}{2015-Augus}. \bibinfo{pages}{146--156}.
\newblock
\showISBNx{9780769555942}
\showISSN{21601860}
\urldef\tempurl%
\url{https://doi.org/10.1109/MSR.2015.21}
\showDOI{\tempurl}


\bibitem[\protect\citeauthoryear{Casteigts, Flocchini, Quattrociocchi, and
  Santoro}{Casteigts et~al\mbox{.}}{2012}]%
        {Casteigts2012}
\bibfield{author}{\bibinfo{person}{Arnaud Casteigts}, \bibinfo{person}{Paola
  Flocchini}, \bibinfo{person}{Walter Quattrociocchi}, {and}
  \bibinfo{person}{Nicola Santoro}.} \bibinfo{year}{2012}\natexlab{}.
\newblock \showarticletitle{Time-varying graphs and dynamic networks}.
\newblock \bibinfo{journal}{\emph{International Journal of Parallel, Emergent
  and Distributed Systems}} \bibinfo{volume}{27}, \bibinfo{number}{5}
  (\bibinfo{date}{oct} \bibinfo{year}{2012}), \bibinfo{pages}{387--408}.
\newblock
\showISSN{1744-5760}
\urldef\tempurl%
\url{https://doi.org/10.1080/17445760.2012.668546}
\showDOI{\tempurl}


\bibitem[\protect\citeauthoryear{de~Fran{\c{c}}a and Ali}{de~Fran{\c{c}}a and
  Ali}{2020}]%
        {DeFranca2020}
\bibfield{author}{\bibinfo{person}{Breno Bernard~Nicolau de Fran{\c{c}}a} {and}
  \bibinfo{person}{Nauman~Bin Ali}.} \bibinfo{year}{2020}\natexlab{}.
\newblock \showarticletitle{The Role of Simulation-Based Studies in Software
  Engineering Research}.
\newblock In \bibinfo{booktitle}{\emph{Contemporary Empirical Methods in
  Software Engineering}}. \bibinfo{publisher}{Springer International
  Publishing}, \bibinfo{address}{Cham}, \bibinfo{pages}{263--287}.
\newblock
\showISBNx{9783030324896}
\urldef\tempurl%
\url{https://doi.org/10.1007/978-3-030-32489-6_10}
\showDOI{\tempurl}


\bibitem[\protect\citeauthoryear{de~Fran{\c{c}}a and Travassos}{de~Fran{\c{c}}a
  and Travassos}{2016}]%
        {DeFranca2016}
\bibfield{author}{\bibinfo{person}{Breno Bernard~Nicolau de Fran{\c{c}}a} {and}
  \bibinfo{person}{Guilherme~Horta Travassos}.}
  \bibinfo{year}{2016}\natexlab{}.
\newblock \showarticletitle{Experimentation with dynamic simulation models in
  software engineering: planning and reporting guidelines}.
\newblock \bibinfo{journal}{\emph{Empirical Software Engineering}}
  \bibinfo{volume}{21}, \bibinfo{number}{3} (\bibinfo{date}{jun}
  \bibinfo{year}{2016}), \bibinfo{pages}{1302--1345}.
\newblock
\showISSN{1382-3256}
\urldef\tempurl%
\url{https://doi.org/10.1007/s10664-015-9386-4}
\showDOI{\tempurl}


\bibitem[\protect\citeauthoryear{Dorner}{Dorner}{2022a}]%
        {Dorner2022software}
\bibfield{author}{\bibinfo{person}{Michael Dorner}.}
  \bibinfo{year}{2022}\natexlab{a}.
\newblock \bibinfo{booktitle}{\emph{michaeldorner/only-time-will-tell: v2.0}}.
\newblock
\urldef\tempurl%
\url{https://doi.org/10.5281/zenodo.6719261}
\showDOI{\tempurl}


\bibitem[\protect\citeauthoryear{Dorner}{Dorner}{2022b}]%
        {Dorner2022data}
\bibfield{author}{\bibinfo{person}{Michael Dorner}.}
  \bibinfo{year}{2022}\natexlab{b}.
\newblock \bibinfo{booktitle}{\emph{Only Time Will Tell}}.
\newblock
\urldef\tempurl%
\url{https://doi.org/10.5281/zenodo.6542540}
\showDOI{\tempurl}


\bibitem[\protect\citeauthoryear{Felderer, Horta, and Editors}{Felderer
  et~al\mbox{.}}{2020}]%
        {Felderer2020}
\bibfield{author}{\bibinfo{person}{Michael Felderer},
  \bibinfo{person}{Guilherme Horta}, {and} \bibinfo{person}{Travassos
  Editors}.} \bibinfo{year}{2020}\natexlab{}.
\newblock \bibinfo{booktitle}{\emph{Contemporary Empirical Methods in Software
  Engineering}}.
\newblock \bibinfo{publisher}{Springer International Publishing}.
\newblock
\showISBNx{978-3-030-32488-9}
\urldef\tempurl%
\url{https://doi.org/10.1007/978-3-030-32489-6}
\showDOI{\tempurl}


\bibitem[\protect\citeauthoryear{Gote, Scholtes, and Schweitzer}{Gote
  et~al\mbox{.}}{2021}]%
        {Gote2021}
\bibfield{author}{\bibinfo{person}{Christoph Gote}, \bibinfo{person}{Ingo
  Scholtes}, {and} \bibinfo{person}{Frank Schweitzer}.}
  \bibinfo{year}{2021}\natexlab{}.
\newblock \showarticletitle{Analysing Time-Stamped Co-Editing Networks in
  Software Development Teams using git2net}.
\newblock \bibinfo{journal}{\emph{Empirical Software Engineering}}
  \bibinfo{volume}{26} (\bibinfo{date}{7} \bibinfo{year}{2021}),
  \bibinfo{pages}{75}.
\newblock
Issue 4.
\showISSN{1382-3256}
\urldef\tempurl%
\url{https://doi.org/10.1007/s10664-020-09928-2}
\showDOI{\tempurl}


\bibitem[\protect\citeauthoryear{Hagberg, Schult, and Swart}{Hagberg
  et~al\mbox{.}}{2008}]%
        {Hagberg2008}
\bibfield{author}{\bibinfo{person}{A~A Hagberg}, \bibinfo{person}{D~A Schult},
  {and} \bibinfo{person}{P~J Swart}.} \bibinfo{year}{2008}\natexlab{}.
\newblock \showarticletitle{Exploring network structure, dynamics, and function
  using NetworkX}.
\newblock \bibinfo{journal}{\emph{7th Python in Science Conference (SciPy
  2008)}} \bibinfo{number}{SciPy} (\bibinfo{year}{2008}),
  \bibinfo{pages}{11--15}.
\newblock


\bibitem[\protect\citeauthoryear{Herbold, Amirfallah, Trautsch, and
  Grabowski}{Herbold et~al\mbox{.}}{2021}]%
        {Herbold2021}
\bibfield{author}{\bibinfo{person}{Steffen Herbold}, \bibinfo{person}{Aynur
  Amirfallah}, \bibinfo{person}{Fabian Trautsch}, {and} \bibinfo{person}{Jens
  Grabowski}.} \bibinfo{year}{2021}\natexlab{}.
\newblock \showarticletitle{A systematic mapping study of developer social
  network research}.
\newblock \bibinfo{journal}{\emph{Journal of Systems and Software}}
  \bibinfo{volume}{171} (\bibinfo{date}{jan} \bibinfo{year}{2021}),
  \bibinfo{pages}{110802}.
\newblock
\showISSN{01641212}
\urldef\tempurl%
\url{https://doi.org/10.1016/j.jss.2020.110802}
\showDOI{\tempurl}


\bibitem[\protect\citeauthoryear{Horta and Barros}{Horta and Barros}{2003}]%
        {Horta2003}
\bibfield{author}{\bibinfo{person}{Guilherme Horta} {and}
  \bibinfo{person}{Travassos Márcio De~Oliveira Barros}.}
  \bibinfo{year}{2003}\natexlab{}.
\newblock \showarticletitle{Contributions of In Virtuo and In Silico
  Experiments for the Future of Empirical Studies in Software Engineering}.
\newblock \bibinfo{journal}{\emph{2nd Workshop on empirical software
  engineering the future of empirical studies in software engineering}},
  \bibinfo{pages}{117--130}.
\newblock


\bibitem[\protect\citeauthoryear{Lamba, Trockman, Armanios, K{\"{a}}stner,
  Miller, and Vasilescu}{Lamba et~al\mbox{.}}{2020}]%
        {Lamba2020}
\bibfield{author}{\bibinfo{person}{Hemank Lamba}, \bibinfo{person}{Asher
  Trockman}, \bibinfo{person}{Daniel Armanios}, \bibinfo{person}{Christian
  K{\"{a}}stner}, \bibinfo{person}{Heather Miller}, {and}
  \bibinfo{person}{Bogdan Vasilescu}.} \bibinfo{year}{2020}\natexlab{}.
\newblock \showarticletitle{Heard it through the Gitvine: an empirical study of
  tool diffusion across the npm ecosystem}. In
  \bibinfo{booktitle}{\emph{Proceedings of the 28th ACM Joint Meeting on
  European Software Engineering Conference and Symposium on the Foundations of
  Software Engineering}}. \bibinfo{publisher}{ACM}, \bibinfo{address}{New York,
  NY, USA}, \bibinfo{pages}{505--517}.
\newblock
\showISBNx{9781450370431}
\urldef\tempurl%
\url{https://doi.org/10.1145/3368089.3409705}
\showDOI{\tempurl}


\bibitem[\protect\citeauthoryear{M{\"{u}}ller and Pfahl}{M{\"{u}}ller and
  Pfahl}{2008}]%
        {Muller2008}
\bibfield{author}{\bibinfo{person}{Mark M{\"{u}}ller} {and}
  \bibinfo{person}{Dietmar Pfahl}.} \bibinfo{year}{2008}\natexlab{}.
\newblock \showarticletitle{Simulation Methods}.
\newblock In \bibinfo{booktitle}{\emph{Guide to Advanced Empirical Software
  Engineering}}. \bibinfo{publisher}{Springer London},
  \bibinfo{address}{London}, \bibinfo{pages}{117--152}.
\newblock
\showISBNx{9781848000438}
\urldef\tempurl%
\url{https://doi.org/10.1007/978-1-84800-044-5_5}
\showDOI{\tempurl}


\bibitem[\protect\citeauthoryear{Nia, Bird, Devanbu, and Filkov}{Nia
  et~al\mbox{.}}{2010}]%
        {Nia2010}
\bibfield{author}{\bibinfo{person}{Roozbeh Nia}, \bibinfo{person}{Christian
  Bird}, \bibinfo{person}{Premkumar Devanbu}, {and} \bibinfo{person}{Vladimir
  Filkov}.} \bibinfo{year}{2010}\natexlab{}.
\newblock \showarticletitle{Validity of network analyses in Open Source
  Projects}. In \bibinfo{booktitle}{\emph{2010 7th IEEE Working Conference on
  Mining Software Repositories (MSR 2010)}}. \bibinfo{publisher}{IEEE},
  \bibinfo{pages}{201--209}.
\newblock
\showISBNx{978-1-4244-6802-7}
\showISSN{02705257}
\urldef\tempurl%
\url{https://doi.org/10.1109/MSR.2010.5463342}
\showDOI{\tempurl}


\bibitem[\protect\citeauthoryear{Nicosia, Tang, Musolesi, Russo, Mascolo, and
  Latora}{Nicosia et~al\mbox{.}}{2012}]%
        {Nicosia2012}
\bibfield{author}{\bibinfo{person}{Vincenzo Nicosia}, \bibinfo{person}{John
  Tang}, \bibinfo{person}{Mirco Musolesi}, \bibinfo{person}{Giovanni Russo},
  \bibinfo{person}{Cecilia Mascolo}, {and} \bibinfo{person}{Vito Latora}.}
  \bibinfo{year}{2012}\natexlab{}.
\newblock \showarticletitle{Components in time-varying graphs}.
\newblock \bibinfo{journal}{\emph{Chaos: An Interdisciplinary Journal of
  Nonlinear Science}} \bibinfo{volume}{22}, \bibinfo{number}{2}
  (\bibinfo{date}{jun} \bibinfo{year}{2012}), \bibinfo{pages}{023101}.
\newblock
\showISSN{1054-1500}
\urldef\tempurl%
\url{https://doi.org/10.1063/1.3697996}
\showDOI{\tempurl}


\bibitem[\protect\citeauthoryear{Ouvrard}{Ouvrard}{2020}]%
        {Ouvrard2020}
\bibfield{author}{\bibinfo{person}{Xavier Ouvrard}.}
  \bibinfo{year}{2020}\natexlab{}.
\newblock \showarticletitle{Hypergraphs: an introduction and review}.
\newblock  (\bibinfo{date}{feb} \bibinfo{year}{2020}).
\newblock


\bibitem[\protect\citeauthoryear{Robinson}{Robinson}{2006}]%
        {Robinson2006}
\bibfield{author}{\bibinfo{person}{Stewart Robinson}.}
  \bibinfo{year}{2006}\natexlab{}.
\newblock \showarticletitle{Conceptual Modeling for Simulation: Issues and
  Research Requirements}. In \bibinfo{booktitle}{\emph{Proceedings of the 2006
  Winter Simulation Conference}}. \bibinfo{publisher}{IEEE},
  \bibinfo{pages}{792--800}.
\newblock
\showISBNx{1-4244-0501-7}
\showISSN{08917736}
\urldef\tempurl%
\url{https://doi.org/10.1109/WSC.2006.323160}
\showDOI{\tempurl}


\bibitem[\protect\citeauthoryear{Rogers}{Rogers}{2003}]%
        {Rogers2003}
\bibfield{author}{\bibinfo{person}{Everett~M. Rogers}.}
  \bibinfo{year}{2003}\natexlab{}.
\newblock \bibinfo{booktitle}{\emph{Diffusion of Innovations}}.
\newblock \bibinfo{publisher}{Free Press}, \bibinfo{address}{New York}. 576
  pages.
\newblock
\showISBNx{9780743258234}


\end{thebibliography}
